\documentclass[aps,prl,twocolumn,superscriptaddress,
,floatfix]{revtex4}
\usepackage{graphicx}
\usepackage{amsmath,amssymb}

\begin{document}

\title{Entangled photon pairs from quantum dots embedded in a self-aligned cavity}

\author{Laia Gin\'{e}s}
\affiliation{Technische Physik, Physikalisches Institut and W\"{u}rzburg-Dresden Cluster of Excellence ct.qmat,Universit\"{a}t W\"{u}rzburg, Am Hubland, D-97074 W\"{u}rzburg, Germany}
\email[]{laia.gines@fysik.su.se}
\author{Carlo Pepe}
\affiliation{Department of Physics, Stockholm University, 10691 Stockholm, Sweden}
\author{Junior Gonzales}
\affiliation{Department of Physics, Stockholm University, 10691 Stockholm, Sweden}
\author{Niels Gregersen}
\affiliation{DTU Fotonik, Department of Photonics Engineering, Technical University of Denmark, Building 343, DK-2800 Kongens Lyngby, Denmark}
\author{Sven H\"{o}fling}
\affiliation{Technische Physik, Physikalisches Institut and W\"{u}rzburg-Dresden Cluster of Excellence ct.qmat,Universit\"{a}t W\"{u}rzburg, Am Hubland, D-97074 W\"{u}rzburg, Germany}
\author{Christian Schneider}
\affiliation{Technische Physik, Physikalisches Institut and W\"{u}rzburg-Dresden Cluster of Excellence ct.qmat,Universit\"{a}t W\"{u}rzburg, Am Hubland, D-97074 W\"{u}rzburg, Germany}
\affiliation{Institute of Physics, University of Oldenburg, D-26129 Oldenburg, Germany}
\author{Ana Predojevi\'{c}}
\email[]{ana.predojevic@fysik.su.se}
\affiliation{Department of Physics, Stockholm University, 10691 Stockholm, Sweden}

\begin{abstract}
We introduce a scalable photonic platform that enables efficient generation of entangled photon pairs from a semiconductor quantum dot. Our system, which is based on a self-aligned quantum dot- micro-cavity structure, erases the need for complex steps of lithography and nanofabrication. We experimentally show collection efficiency of 0.17 combined with a Purcell enhancement of up to 1.7 in the pair emission process. We harness the potential of our device to generate  photon pairs entangled in time bin, reaching a fidelity of 0.84(5) with the maximally entangled state. The achieved pair collection efficiency is 4 times larger than the state-of-the art. The device, which theoretically supports pair extraction efficiencies of nearly 0.5 is a promising candidate for the implementation of bright sources of time-bin, polarization- and hyper entangled photon pairs in a straightforward manner. 
\end{abstract}


\maketitle

Photon entanglement is an essential element of numerous quantum communication protocols \cite{Horodecki4}. Entangled photons can be generated by different processes including spontaneous parametric downconversion, decay of atomic systems, and recombination of biexciton in quantum dots. While parametric downconversion is still the approach that achieves the highest values of entanglement \cite{Fedrizzi,Sagnac}, the recent developments of quantum dot devices promise to provide sources with a comparable quality \cite{DHuber18}. Furthermore, quantum dots as photon sources allow for sub-Poissonian statistics, which is an additional asset for establishing high rate and safe quantum communication \cite{Hosak20}. The entanglement of photons generated by quantum dots has been shown in polarization \cite{akopian2006}, in time bin \cite{time-bin}, and also as hyperentanglement \cite{hyper}. However, the vast majority of results shown up to now have been made employing quantum dot structures that featured very limited extraction efficiency and no Purcell effect. The exemptions to this rule are few including \cite{Dousse2010} and  \cite{bull_pan, liu2019}. Such devices employ an engineered photonic environment that requires to be accurately aligned to the site of the quantum dot formation, in order to exploit cavity quantum electrodynamics and enhance the emission into free space. While there are approaches that allow for aforesaid alignment including site-selective quantum dot growth \cite{schneider2009}, in-situ lithography \cite{Dousse2010}, and  quantum dot spectral imaging \cite{thon2009, Sapienza2015}, they commonly bring along a significant increase in the fabrication complexity. On the other hand, a small mode volume microcavity based on distributed Bragg reflectors can also form naturally in the growth process \cite{Zajac12}, through deformation of the top distributed Bragg reflector (DBR) mirror. This deformation is triggered by a crystal defect in the bottom DBR, which propagates through the entire device. It yields a dimple with a size of a few micrometers that enables the confinement of the optical mode. In structures with embedded quantum dots, the same crystalline defect that triggers the deformation of the DBRs also seeds the quantum dot nucleation, yielding highly performing self-aligned cavity structures. Such structures have been exploited for efficient extraction of single photons \cite{Maier14}, spin-photon interfacing \cite{deGreve}, and single emitter coherent control via resonant four wave mixing \cite{Fras2016}.  While, it was theoretically predicted that self-aligned quantum dot-cavity structures may allow for efficient photon extraction over a broad spectral range \cite{Maier14}, they have not been investigated as structures that support generation of entangled photon pairs.  Here, we generate photon pairs entangled in time bin and we show that the device can be exploited for broadband and efficient extraction of photon pairs.

A planar microcavity does not commonly provide a lateral confinement, and therefore a defined mode volume. However, if a deformation of the top DBR \cite{Maier14}  above the site of quantum dot formation occurs, the resulting cavity will not act as planar anymore and it will allow for lateral confinement. Consequently, effects such as shortening of the lifetime of the embedded emitters and an increase in the collection efficiency should be observed. To characterize the emitters in our sample we performed several measurements including lifetime, auto, and cross-correlation. 

\begin{figure}[ht]
\centering\includegraphics[width=0.70\linewidth]{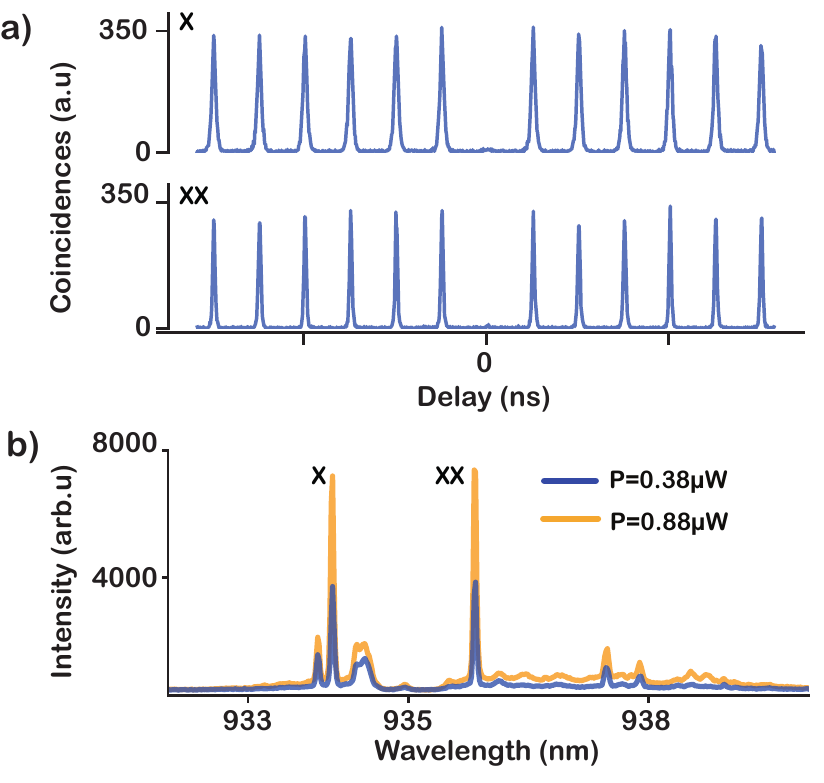}
\caption{ a) The autocorrelation measurements of the exciton and biexciton photons  b) Emission spectrum for two different excitation powers}
\label{hom1}
\end{figure}

The autocorrelation measurements performed revealed high purity of the emitted state yielding  $g^{(2)}(0)_{x}$=0.025(3) and $g^{(2)}(0)_{xx}$=0.016(3), for exciton and biexciton, respectively.  The results of the autocorrelation measurements are shown in Fig.\ref{hom1}a. The sample was kept for measurements in a helium-flow cryostat stabilized to 4.0 $\pm$ 0.05 K. The quantum dots were driven resonantly using two-photon resonant excitation of the biexciton \cite{twophoton}. Fig.\ref{hom1}b shows the emission spectra of a quantum dot emitter for two different excitation powers, 0.38$\mu W$ and 0.88$\mu W$. The excitation pulses were derived from a 80MHz Ti:Sapphire laser. The length of the excitation pulses was adjusted by means of a pulse-stretcher, built in a 4f configuration \cite{twophoton}. The emission and excitation paths were co-linear and the excess laser scattering was filtered by means of a polarizer and a pair of notch filters with a bandwidth of 0.5~nm. The quantum dot emission (biexciton and exciton photons) was spectrally separated using a home-built spectrometer \cite{twophoton} and coupled into single mode fibers. 

\begin{figure}[t]
\centering\includegraphics[width=0.68\linewidth]{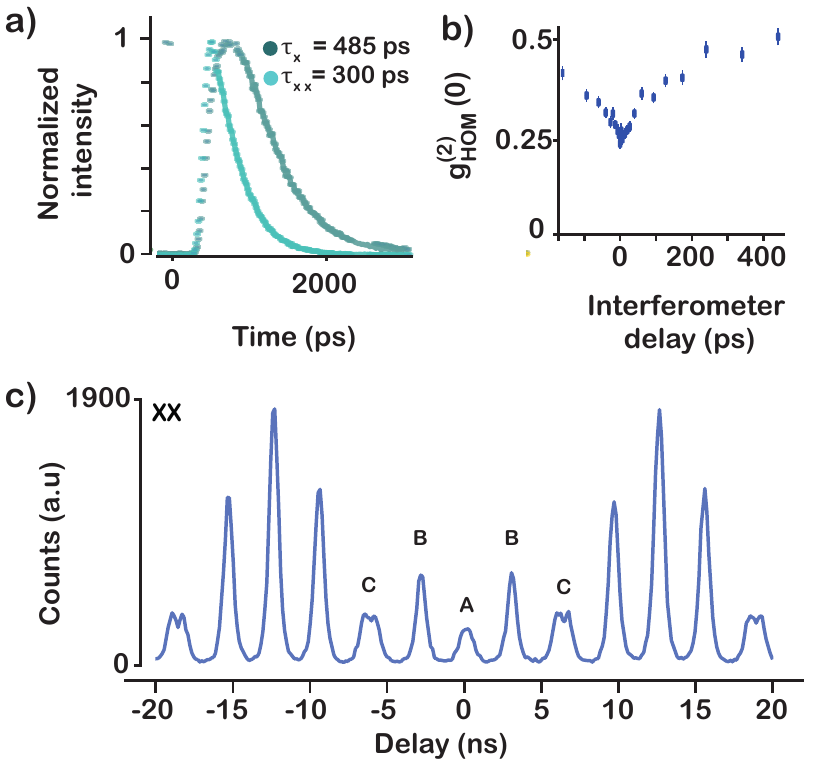}
\caption{ a) Results of the lifetime measurements for exciton and biexciton b) Two photon interference correlation as a function of the delay introduced by changing the interferometer length  c) Correlation function $g^{(2)}_{xxHOM}(\tau)$, obtained in Hong-Ou-Mandel interference measurement of consecutively emitted photons at zero delay performed using biexciton photons.}
\label{hom2}
\end{figure}

The lifetime measurements were recorded using a single photon detector with 16~ps resolution. The results are shown in Fig.\ref{hom2}a. To prove the reproducible nature of our approach we performed the lifetime measurements on several quantum dot emitters. They showed similar lifetime values ranging $\tau_{x}$=~468(8)-540(7)~ps, and   $\tau_{xx}$=~300(3)-334(3)~ps, for exciton and biexciton, respectively. The values we obtained indicate a moderate Purcell enhancement (ranging from 1.2 to 1.7). This estimate was made considering an average value of $\tau_{xx0}$=400~ps and $\tau_{x0}$=800~ps, for biexciton and exciton photon respectively, when quantum dots are not cavity embedded.

We characterized the indistinguishibility of the consecutively emitted photons by exploiting Hong-Ou-Mandel interference. We have implemented this measurement using two Michelson interferometers with a nominal delay of 3~ns. The first interferometer served to generate two excitation laser pulses sent to excite the quantum dot, while the second one was used to observe interference of single photons \cite{HOM}, emitted in two consecutive excitations. The delay of one of the interferometers was adjustable such that the arrival time of the photons could be changed to make them distinguishable, as shown in Fig.\ref{hom2}b.  The results obtained for biexciton photons emitted at zero delay at the beamsplitter are shown in Fig.\ref{hom2}c. The five peaks in this type of measurement (A-C) arise from three different types of coincidence events. Peak A represents the simultaneous arrival of the two photons at the beamsplitter. Here, the photon created by the first (second) excitation pulse travels along the long (short) path on the interferometer, respectively. Registered coincidences in peak B ($\pm$ 3~ns) are a result of photons taking the same path (either short or long), whereas peak C ($\pm$ 6~ns) represents the coincidences between the first photon following the short path and the second photon following the long one. From the data and following \cite{Santori}, we get a $g^{(2)}_{xxHOM}(0)$ = 0.259(6), probing biexciton photons indistinguishability. Additionally, the measured exciton photon indistinguishability resulted in $g^{(2)}_{xHOM}(0) $= 0.286(23). The corresponding two photon interference visibility values were measured to be 0.508(10) for biexciton and 0.44(4) for exciton.

\begin{figure}[t]
\centering\includegraphics[width=1\linewidth]{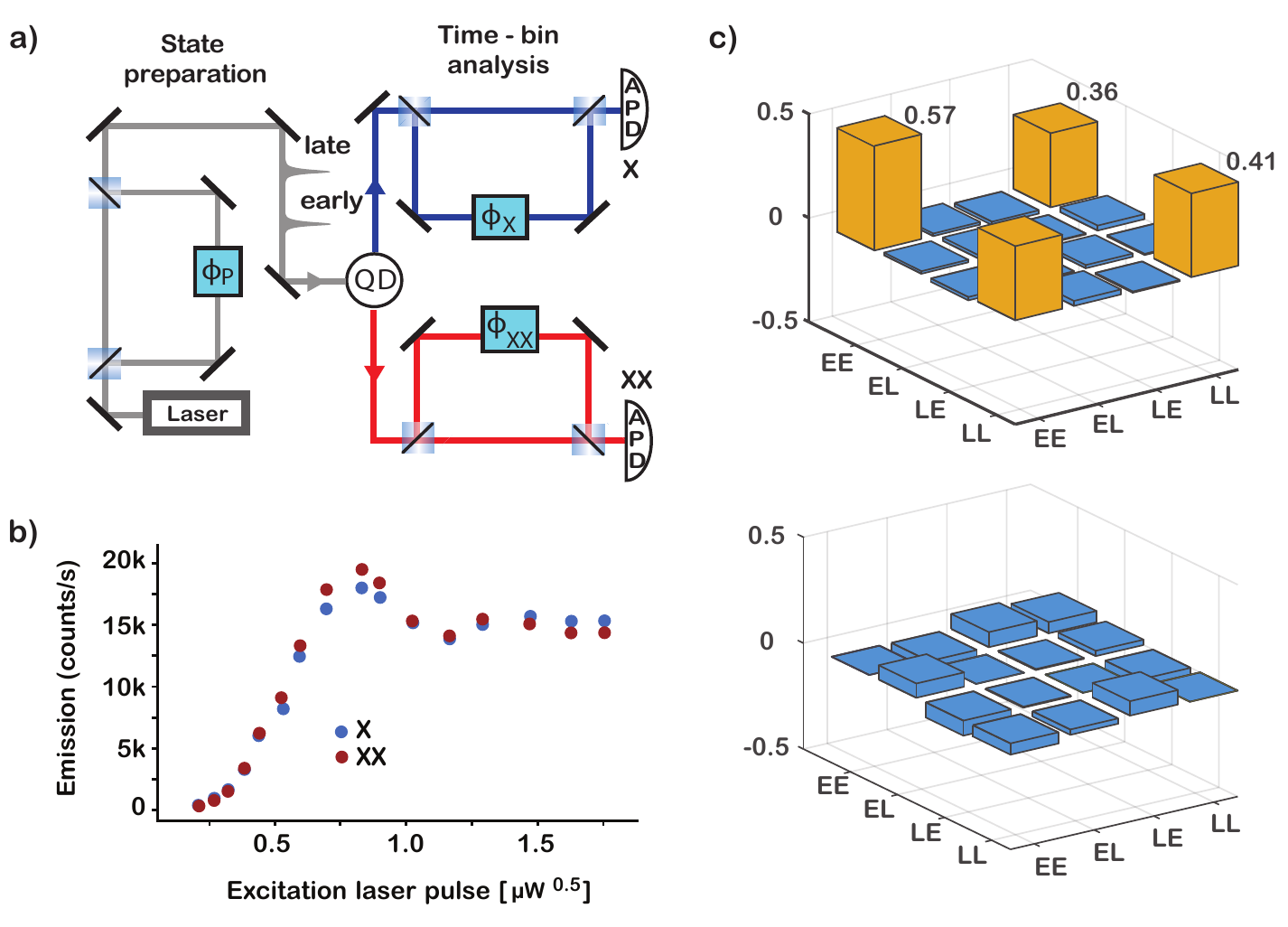}
\caption{ a) The scheme of the experimental set-up used for generation and measurement of the time-bin entanglement. The quantum dot system is excited by two consecutive pulses obtained from an unbalanced interferometer shown on the left. The relative phase between these pulses is $\phi_P$. The state analysis is performed using two additional interferometers, one for the exciton photon and the other for the biexciton photon. the phases of these to interferometers are $\phi_X$ and $\phi_{XX}$, respectively. The photons are detected upon leaving the analysis interferometers using avalanche photo detectors. b) Exciton and biexciton Rabi oscillations as a function of the excitation laser pulse area c) Real and imaginary part of the measured density matrix.}
\label{matrix}
\end{figure}

The time-bin entanglement was generated and characterized using an optical system consisting of three mutually phase stable Michelson interferometers \cite{time-bin}. This type of entanglement encodes the state in a superposition of the system's excitation times or time bins named \textit{early} and \textit{late}. Therefore, one of the three interferometers, termed pump interferometer, was used to generate the early and late laser pulses exciting the quantum dot (shown in Fig.\ref{matrix}a). The remaining two interferometers were used to analyse the entanglement \cite{brendel, time-bin}. The delay in the interferometers was set to be 3~ns, which is longer than the coherence of the photons emitted by the quantum dot.  The relative phases between the pump and the analysis interferometers were adjusted by means of phase plates placed in individual interferometers \cite{time-bin}.

To generate the time-bin entanglement we require to prepare the system in a superposition of being excited by the early or by the late pulse. The phase of the superposition is determined by the phase of the pump interferometer. We transfer the phase $\phi_P$ onto the system by driving it resonantly and coherently. We do so by employing the two-photon resonant excitation. The resonant nature of the excitation can be confirmed by Rabi oscillations shown in Fig.\ref{matrix}b. The entangled state was characterized by means of state tomography using 16 projective measurements \cite{james, takesue}. The density matrix of the entangled state is shown in Fig.\ref{matrix}c. It yields a concurrence of 0.70(10) and fidelity to the maximally entangled state of 0.84(5). In order to obtain the measurement errors we performed a 50 run Monte Carlo simulation of the data with a Poissonian noise model applied to the measured values.

The resonant excitation allows us to accurately estimate the photon collection efficiency in the first lens. Under the laser excitation rate of 80~MHz and using a pulse area that maximizes the emission probability, we observed the count rate of 61kcounts/s and 26kcounts/s 
for biexciton and exciton 
photons, respectively.  
This result was achieved using detectors with quantum efficiency of $0.25$, the efficiency of coupling into a single mode fiber $0.4$ 
for biexciton and $0.18$ for exciton photon, 
and an overall optical setup efficiency of $0.12$. The effective excitation rate was reduced due to blinking to 0.625 of the nominal value. This number was obtained by comparing the number of coincidence event at short and long delay times in an autocorrelation measurement. The emission probability estimated from Rabi oscillations was 0.65. These numbers yield efficiency of $0.17$ of collection in the first lens above the sample. The lens used had a numerical aperture of 0.62.

Our sample contained self-assembled InAs quantum dots grown by molecular beam epitaxy. The In(Ga)As quantum dots were grown via indium flush technique and embedded between 24(5) bottom(top) AlGaAs/GaAs mirror pairs that form a micro cavity of thickness $\lambda$ and with a mode at $\lambda$ =936~nm. The development of similar structures has been analyzed in \cite{schneider2015single, Maier14}. The structure layer layout was chosen to match the emission of the quantum dots with the resonance of the cavity. 

\begin{figure}[ht]
\centering\includegraphics[width=0.65\linewidth]{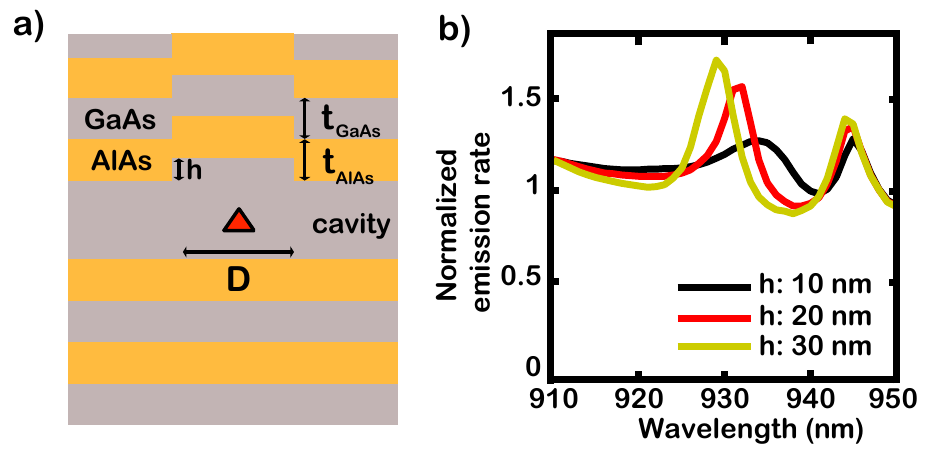}
\caption{a) An illustration of the simulated geometry b) Purcell effect as a function of wavelength for h=10~nm (black), h=20~nm (red) and h=30~nm (yellow). Parameters of the  simulation are: $t_{GaAs}$=68~nm, $t_{AlAs}$=82~nm, $t_{cavity}$=270~nm and D=2000~nm.}
\label{sim}
\end{figure}

We performed numerical simulations of the Purcell factor using an eigenmode expansion technique  \cite{Hayrynen}. The system was modelled as a cavity featuring a quantum dot in its center and a conical circular defect above the quantum dot as illustrated in Fig.\ref{sim}a. The computed Purcell factor is presented in Fig.\ref{sim}b as function of wavelength and defect height h. As compared to the planar case, the defect also has a role of a lens, finally enhancing \cite{Maier14} the collection efficiency to nearly 0.5 at the cavity resonance for h=20~nm and NA=0.7.  We account that the use of a lower NA collection lens (0.62) is in part responsible for the reduced experimentally observed efficiency compared to the theory. On the other hand, the Purcell enhancement we observe is in good agreement with the model.

The further development of quantum technologies and quantum communication enforces a search for more efficient and better performing quantum sources and devices. Systems consisting of quantum dots embedded in micro cavities hold a promise to provide us with high efficiency and rate entangled photons on-demand. However, the current attempts to maximize the extraction efficiency involve sophisticated engineered photonic systems that call for accurate alignment with the emitter. All of this implies the need of an intricate nanofabrication processes.
In this work we presented a simple, scalable photonic device that grants efficient collection of entangled photon pairs from InAs quantum dots. The device performance was thoroughly characterized. We showed experimental collection efficiencies of 0.17, and a Purcell of up to 1.7, of photons entangled in time-bin. Our result yields collection efficiency of time-bin entangled photons that is 4 times greater than achieved up to date. The achieved entanglement was characterized and we showed a concurrence of 0.70(10) and a fidelity of 0.84(5). 
The performance of the device can be further enhanced by equipping it with a strain actuator \cite{DHuber18}. This step would further enhance the versatility of the device by allowing also for the generation of polarization and hyperentanglement.

\begin{acknowledgments}
We acknowledge funding by the DFG within the projects SCHN1376-5.1 and PR1749/1-1. A.P. would like to acknowledge Swedish Research Council and Carl Tryggers Stiftelse.
\end{acknowledgments}

\end{document}